\documentclass{jpsj-suppl}
\usepackage{times}
\usepackage{color}
\def\prb{Phys.\ Rev.\ B }
\def\prl{Phys.\ Rev.\ Lett.\ }

\def\JPCM{J.\ Phys.\ Condens.\ Matter }

\def\RMP{Rev.\ Mod.\ Phys.\ }

\def\ufc{U_{\scriptscriptstyle{fc}}}
\def\ufcc{U^{\scriptscriptstyle C}_{\scriptscriptstyle{fc}}}

\title{Dynamics of valence fluctuations in the extended periodic Anderson model}

\author{Pramod Kumar \thanks{E-mail address: pramod@jncasr.ac.in} and N.\ S.\ Vidhyadhiraja\thanks{E-mail address: raja@jncasr.ac.in} %\\
}
\inst{Theoretical Sciences Unit, 
Jawaharlal Nehru Center for Advanced Scientific Research,\\
Jakkur, Bengaluru 560064, India. 
}
\abst{We have investigated the valence transition in the extended periodic Anderson model (EPAM) within the
framework of dynamical mean field theory. Previous theoretical studies have indicated the presence
of this transition in parameter ranges that are experimentally inaccessible. In this work,
we show that, inclusion of dynamics beyond static mean-field brings the valence transition in physically relevant regimes. 
We  show the strong dependence of the quasiparticle-weight scale
across the sharp valence crossover. The density of states also exhibit concomitant 
large scale spectral weight transfers, which can be observed in experiments. 
}

\kword{heavy fermions, valence fluctuations, quantum criticality, dynamical mean field
theory}

\begin{document}
\maketitle

\section{Introduction}
In recent years, quantum critical phenomena have been one of the focal points for the condensed matter community. Many rare earth inter-metallic compounds, which are also heavy fermions can be tuned easily to a quantum critical point by application of external perturbations like magnetic field and pressure~\cite{gegen}. YbRh$_2$Si$_2$~\cite{cust} and CeCu$_2$Si$_2$~\cite{yuan} are a few examples. The periodic Anderson model (PAM) is the paradigm for studying these kind of systems. There are two possible theoretical scenarios for achieving quantum phase transitions in heavy fermion alloys ~\cite{piers}, namely (a) competition between Kondo interaction (which is local in nature) and the non-local RKKY interactions and (b) Kondo destruction (KD) leading to sharp valence transition and consequently a quantum phase transition. However, it is established
through numerous studies that within the PAM, while a smooth valence crossover is possible, there is no possibility of a valence transition for a smooth change in model parameters.

     The extended periodic Anderson model (EPAM)~\cite{miyake}, which includes a Coulomb repulsion ($\ufc$) between conduction ($c$) and localised electrons ($f$) naturally captures the second scenario and has been studied using Hartree-Fock mean field theory~\cite{hews}, slave-boson (SB) mean-field theory ~\cite{miyake} and dynamical mean-field theory (DMFT) using exact diagonalization (ED)~\cite{sugi} as the impurity solver. Nevertheless, the model parameters reported for the
occurrence of the transition is inaccessible physically. Recently, a continuous time quantum Monte Carlo within DMFT~\cite{yoshida} study of the broken symmetry phases in the particle-hole symmetric  EPAM was also carried out, but the valence transition was not studied. 

   We have employed a local moment approach (LMA)~\cite{loga98} within DMFT~\cite{geor,prus95}. The LMA has
 been extensively benchmarked for the single-impurity Anderson model against numerical renormalization group and Bethe Ansatz results~\cite{bull}. The agreement has been excellent. The LMA+DMFT investigations of the PAM have also been very successful in describing the spectral and 
transport properties of a large range of heavy fermion~\cite{vidh05,vidh06} and mixed-valent materials~\cite{pram}. In this work, we have investigated the EPAM using LMA. We find that in specific parameter regimes of the EPAM, a quantum critical point (QCP) governed by valence fluctuations exists. Additionally, this parameter regime in which we get the signature of transition is physically accessible. We have reported  the behaviour of low energy scale  and density of states across the valence transition. The large spectral weight transfers found in dynamical quantities should be observable in photoemission experiments.

\section{Model and formalism}
The EPAM  extends the usual PAM by considering an additional  Coulomb interaction between localised electrons and conduction electrons. In the standard second quantised notation, the model may be represented by the following Hamiltonian:
\begin{align}
\hat{H} =  &\epsilon_c \sum_{i\sigma} c_{i\sigma}^\dag c_{i\sigma}^{\phantom{\dag}} -t\sum_{(ij),\sigma} c_{i\sigma}^\dag c_{j\sigma}^{\phantom{\dag}} 
 +V \sum_{i\sigma}(f_{i\sigma}^\dag c_{i\sigma}^{\phantom{\dag}} +{\rm h.c.}) \nonumber\\
 & +\sum_{i\sigma}(\epsilon_f +\frac{U_{f}}{2} f_{i{\bar{\sigma}}}^\dag f_{i{\bar{\sigma}}}^{\phantom{\dag}} )f_{i\sigma}^\dag f_{i\sigma}^{\phantom{\dag}}  + \ufc
 \sum_{i\,\sigma \sigma^{\prime}}n_{i\sigma}^c n_{i\sigma^{\prime}}^f 
\end{align}
The first two terms correspond to the site and kinetic energy of the conduction electrons.    
 The third term hybridises the local ($f$) and itinerant  ($c$) electrons. The fourth term
  intra-$f$-orbital Coulomb repulsion. The last term represents Coulomb repulsion of $c$- and $f$- electrons. Within DMFT~\cite{geor,prus95}, which is exact in the limit of infinite dimensions $D$, the hopping $t_{ij}$ must be rescaled as $t_{ij}\propto t_*/\sqrt{D}$. Decoupling the inter-orbital interaction term within Hartree approximation renormalizes
the  conduction and localised electron site energy as $\tilde{\epsilon_c}=\epsilon_c+\ufc\langle n_f \rangle$ and  $\tilde{\epsilon_f}=\epsilon_f+\ufc\langle n_c \rangle$ where $\langle n_c \rangle$ and $\langle n_f \rangle$ are average occupancies for conduction and localised orbitals respectively. A point to note is that the occupancies, $\langle n_c\rangle $ and 
$\langle n_f\rangle $ are derived from the
full interacting Green's function, and hence must be obtained self-consistently. In this
sense, the present approximation is equivalent to carrying out a skeleton expansion
of the $\ufc$ term about the PAM upto the first order.
 
The retarded Green's functions within DMFT for the localised and conduction electrons
are given respectively by~\cite{pram} 
\begin{eqnarray}
G^f(\omega)=\left[\omega^+ -\tilde{\epsilon_f} - \Sigma(\omega) - \frac{V^2}{\omega^+ -
\tilde{\epsilon_c} - S(\omega)}\right]^{-1} \\
G^c(\omega)=\left[\omega^+ - \tilde{\epsilon_c} - S(\omega) - \frac{V^2}{\omega^+ -
\tilde{\epsilon_f} - \Sigma(\omega)} \right]^{-1}
\end{eqnarray}
where $\Sigma(\omega)$ is the momentum independent $f$-self-energy. The spectral
functions are given by $D^{\alpha}(\omega)=-{\rm Im}G^\alpha(\omega)/\pi$
and the occupancies are given by $n_\alpha=2\int^0_{-\infty}d\omega\,D^\alpha(\omega)$.
 $S(\omega)$
is the Feenberg self-energy, that  may be calculated self-consistently through following relations ~\cite{pram} )
\begin{eqnarray}
G^c(\omega)= H[\gamma] = \int^\infty_{-\infty} \frac{\rho_0(\epsilon)}{
\gamma(\omega)-\epsilon} = \frac{1}{\gamma(\omega)-S(\omega)}
\end{eqnarray}
where $H[\gamma]$ denotes the Hilbert transform of $\gamma$ with respect
to the non-interacting density of states $\rho_0(\epsilon)$. While $\gamma=\omega^+$
in the $V=0$ limit, for finite $V$, it is given by $\gamma(\omega) = 
\omega^+ - \tilde{\epsilon}_c - V^2[\omega^+ - \tilde{\epsilon}_f-\Sigma(\omega)]^{-1}$.

The local $f$-self-energy is calculated using LMA, which has been benchmarked with
 NRG and Bethe ansatz for single impurity Anderson model, with excellent agreement.
The  PAM in various regimes like Kondo Lattice (KL)  and mixed valent (MV) regime has also been studied using LMA. We refer the reader to our earlier works on LMA as applied
to the PAM for more details of the implementation~\cite{pram}. In this work, we have maintained
a constant $n{tot}$ as the $f$-level is varied.

\section{Results and discussion}

\begin{figure}[ht]
\begin{center}
\includegraphics[width=7cm,clip=]{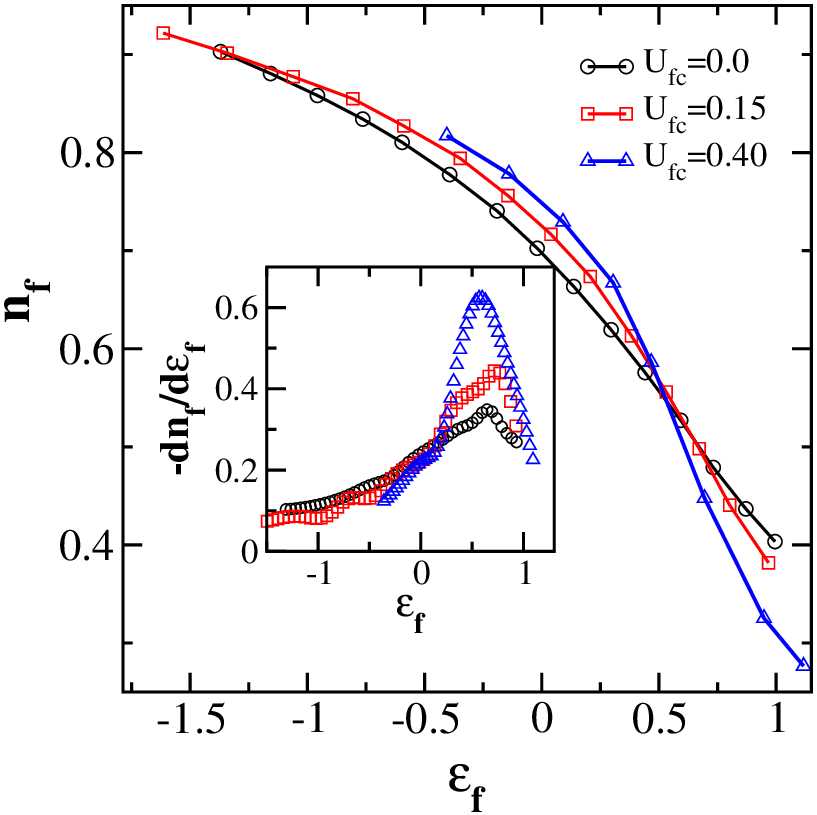}
\end{center}
\caption{(Color online) Variation of $f$-orbital occupancy (main panel) and valence
susceptibility (inset) with  $f$-level, ${\epsilon}_f$, for various values of
$\ufc$. The other 
model parameters are $U\sim 5.3$, $\epsilon_c=0.5$ and $V^2=0.6$.
 }
\label{fig:f1}
\end{figure}

In figure~\ref{fig:f1}, we show the variation of $f$-occupancy ($n_f$) with
the  $f$-level ($\epsilon_f$) energy for various values of $\ufc$. The total
filling $n_{tot}$ is fixed at $\sim 1.5$. It is seen 
that $n_f$ decreases with increasing $ \tilde{\epsilon}_f$ for all $\ufc$. However,
with increasing $\ufc$, the decrease becomes steeper. This is reflected in the
valence susceptibility, shown in the inset of figure~\ref{fig:f1}, which is approaching
a divergent form with increasing $\ufc$. The model parameters (in units of $t_*=1$)
are $U=5.3$, $\epsilon_c=0.5$ and $V^2=0.6$. These numbers for the
parameters are motivated by our earlier works on the PAM~\cite{vidh06}, wherein we have 
successfully compared theory with experiments. Thus, the model parameters
have modest values and are physically accessible.

\begin{figure} [ht]
\begin{center}
\includegraphics[width=8cm,clip=]{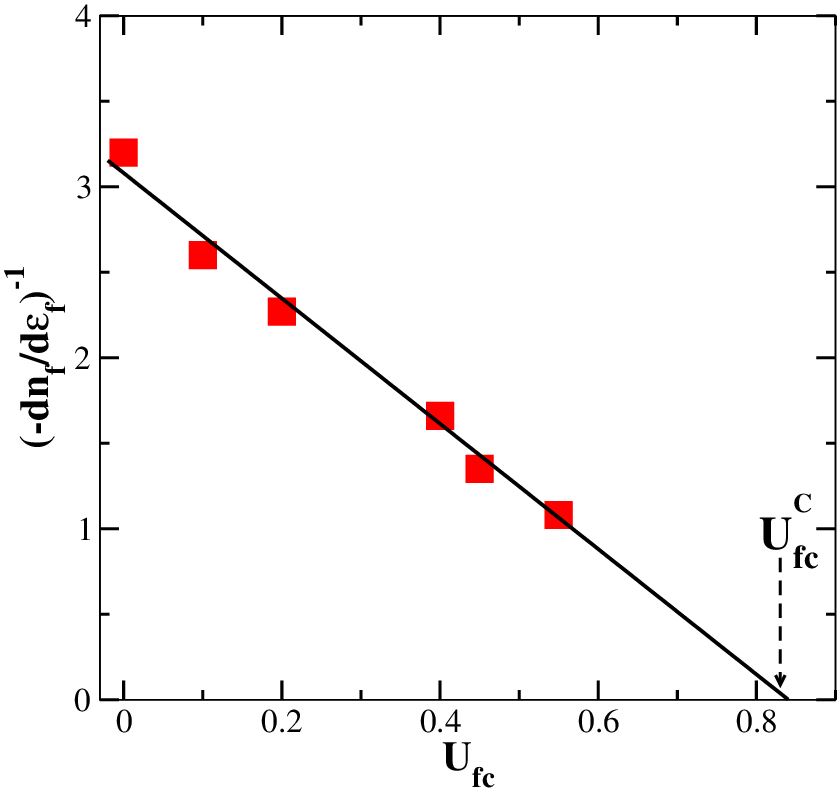}
\end{center}
\caption{(Color online) The black squares are inverse of maximum valence susceptibility
 as a function of $\ufc$. The dashed line is a linear fit extrapolated to the critical value at which the susceptibility would diverge.  The model parameters are 
same as figure ~\ref{fig:f1}
 }
\label{fig:f2}
\end{figure}

The inverse of the maximum susceptibility as a function of $\ufc$,  shown (as filled 
squares)
in figure~\ref{fig:f2}, decreases linearly and, if extrapolated (dashed line), vanishes at 
a critical $\ufcc\sim 0.8$. This is a signature of a valence fluctuations driven quantum critical 
point. Although the same scenario is obtained at static mean field level~\cite{miyake},
the critical range of parameters is so extreme (for $\ufcc\sim 0.8$) , that hardly any experimental
systems would correspond to them. Even in the ED calculations~\cite{sugi}, the parameter regime explored was such that the $U/V^2\sim 500$, because at lower values, the Kondo lattice
regime was found to be unstable. As seen above, we start from very realistic values of
$U, V$ and $\epsilon_c$, and the $\ufcc$ obtained is also a small fraction of the
$U$ scale, and is of the order of the bandwidth ($t_*$), in agreement with previous
theoretical works.
Since the critical valence fluctuations (CVF) model predicts the transition to occur in unreasonably large 
parameter values,
the authors of a recent experimental study on CeCu$_2$Si$_2$~\cite{rauef} propose that the discrepancy between the experimental findings and the predictions of CVF model
could be due to the absence of certain non-local terms in the model. Here, we show that
without the need to include such terms, it is sufficient to include dynamical fluctuations
beyond static mean-field for a realistic and experimentally relevant description of the
pressure-induced valence changes.

\begin{figure} [ht]
\begin{center}
\includegraphics[width=7cm,clip=]{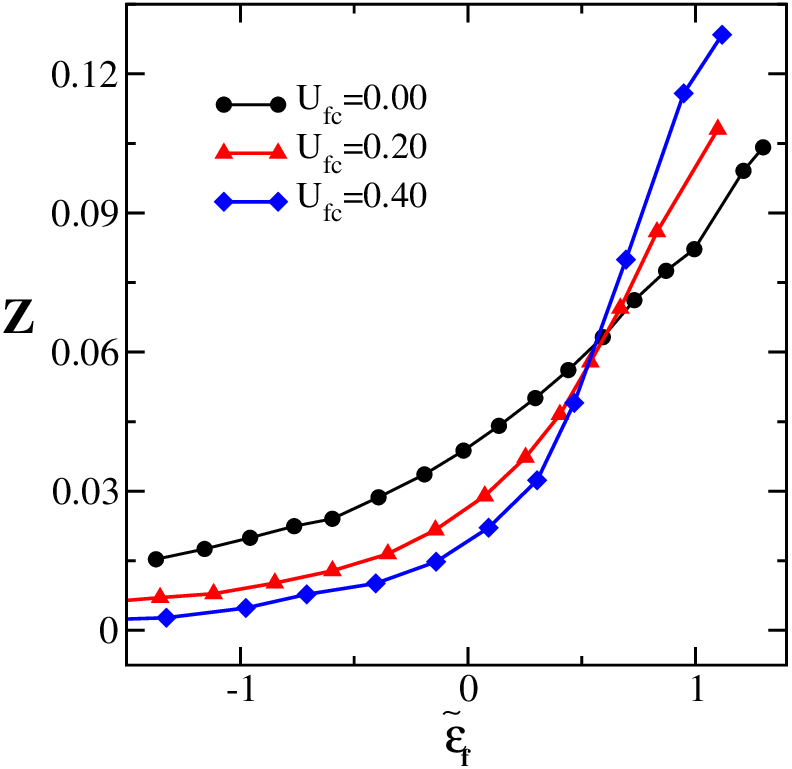}
\end{center}
\caption{(Color online) The quasiparticle weight as a function of the effective 
$f$-level,  $\tilde{\epsilon}_f$ for various $U_{fc}$. The model parameters are same as figure ~\ref{fig:f1}
 }
\label{fig:f3}
\end{figure}

It is important to understand the variation of the low energy coherence scale across this
crossover. As is well-known, the coherence scale is given by $\omega_L=ZV^2/t_*$,
where $Z=(1-\partial\Sigma/\partial\omega)|_{\omega=0}^{-1}$ is the quasiparticle
weight.
In figure ~\ref{fig:f3} we have shown the variation of quasiparticle weight with varying ${\epsilon_f}$ for various values of the inter orbital Coulomb interaction, $\ufc$. The $Z$ increases sharply as the $f$-level moves up in energy.  We observe
another trend for the change in $Z$. For any fixed ${\epsilon_f}$ below the
critical point, the $Z$ increases with increasing $U_{fc}$, while above the critical point,
the low energy scale and $U_{fc}$ are anti-correlated. Thus very close to
 the transition point, the quasiparticle weight is almost independent of changes in $U_{fc}$. It would
be interesting to see such an `isosbestic point' in experiments.

\begin{figure} [ht]
\begin{center}
\includegraphics[width=7cm,clip=]{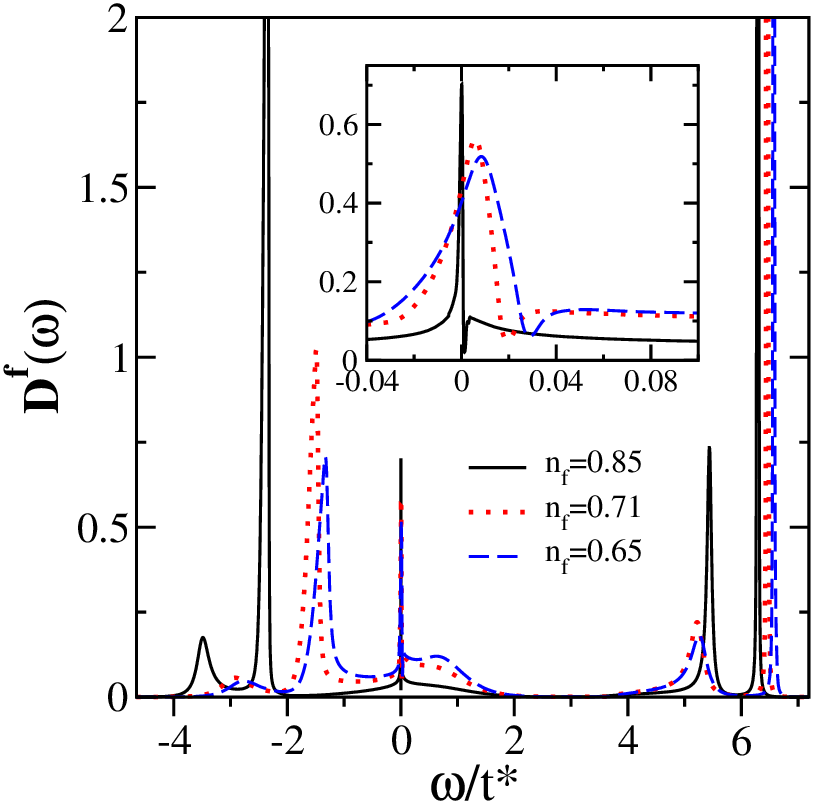}
\end{center}
\caption{(Color online) Main panel:The $f$-density of states as a function of the frequency 
for different $f$-level,  ${\epsilon}_f$ positions, corresponding to
different $n_f $ values, as mentioned in the legends. The $\ufc$ value is fixed at
0.5 and the rest of the model parameters are same as figure~\ref{fig:f1}. Inset: An
expanded view of the spectra in the main panel at low frequencies. 
 }
\label{fig:f4}
\end{figure}

Since the scale and the valence change significantly across the crossover, we must
expect that the density of states (DOS) might exhibit large scale spectral weight transfers.
Indeed, as shown in the main panel of figure~\ref{fig:f4}, the DOS changes significantly across the crossover. With decreasing $n_f$, the lower Hubbard band moves towards
the Fermi level and gets broader, while the upper Hubbard band moves away and gets sharper. As seen from the inset, where an expanded view of the low frequency region
is shown, the Kondo resonance at the Fermi level broadens significantly. This is
 as expected from figure~\ref{fig:f3} where the $Z$ increases with decreasing $n_f$,
and the width of the resonance is proportional to $Z$.

\section{Conclusions}

The extended periodic Anderson model has been proposed as one of the prime
candidate Hamiltonian for describing valence transitions and a new form
of quantum criticality in heavy fermion and mixed-valent systems. However,
the critical point found in theoretical investigations 
lies in model parameter regimes that would not be physically realizable. We present some
preliminary investigations of the EPAM within DMFT using a highly accurate local moment approach as the impurity solver. Our results clearly suggest that the inclusion of
dynamical fluctuations beyond static mean field theories are crucial to making the
critical point 
physically accessible. The low energy Kondo scale changes significantly across the valence
crossover, and the single-particle spectra also show concomitant large scale
weight transfers. These changes must be easily observable in angle
resolved photoemission spectroscopy experiments. Finite temperature investigations
are ongoing and will be important to resolve issues such as the temperature
dependence of the valence changes across the quantum critical point under
the second superconducting dome in CeCu$_2$Si$_2$.
 %\newpage

\end{document}